\def\apj{{ApJ}}

\def\mnras{{MNRAS}}

\documentclass[cup5b]{caps}
\usepackage{graphicx}
\usepackage{amssymb}
\usepackage{ociwsymp1e}
\HeadText{J. D. MacMillan and R. N. Henriksen}

\begin{document}

\pagenumbering{arabic}

\author[]{J. D. MACMILLAN and R. N. HENRIKSEN\\
Department of Physics, Queen's University, Kingston, Ontario, Canada, K7L 3N6}

\chapter{Dynamical Growth of Black Holes \\ 
and the $M_{\rm BH} - \sigma$ Relation}

\begin{abstract}

This poster discusses a possible explanation for the relationship between the mass of the central supermassive black hole and the velocity dispersion in the bulge of the host galaxy.  We suppose that the black hole and the dark matter halo are forming simultaneously as matter falls in, and a self-similar system then exists in which the mass and the velocities of the system evolve as power-law functions of time.  This leads naturally to a relationship between the black hole mass and the velocities in the halo which, with a reasonable choice of cosmological parameters, is in good agreement with the observed relationship.  We also confirm this relationship with more robust numerical results.

\end{abstract}

\section{Dynamical Growth of Black Holes}

Supermassive black holes (BHs) are now considered to be a common feature of galaxies which have a bulge.  Furthermore, a number of observational properties of the host galaxy correlate with the BH mass.  Among the strongest of these correlations is the relationship between the BH mass and velocity dispersion within the galactic bulge (Ferrarese \& Merritt 2000; Gebhardt et al. 2000):  $M_{\rm BH} \propto \sigma^a$, where $a = 4.02 \pm 0.32$ (Tremaine et al. 2002).  Since the velocity dispersion is measured well outside of the BH ``influence,'' this correlation indicates that an intimate relationship exists between the BH and the dynamical structure of the host galaxy.

We assume that a galaxy forms by the extended collapse of a ``halo'' composed of collisionless matter and that simultaneously the central black hole is growing proportionally to the halo as matter continues to fall in.  This is equivalent to the assumption of multidimensional self-similarity (Carter \& Henriksen 1991; Henriksen 1997), and MacMillan \& Henriksen (2002) show that, in this case, the mass $M$ inside any surface in the halo is related to the velocity of the particles by
\begin{equation}
\log{ M}\propto \left(\frac{3\delta/\alpha-2}{\delta/\alpha-1}\right) \log{\sigma},
\label{eq:predict}
\end{equation}
where $\sigma$ is the averaged velocity.  Since the system is self-similar, this relation will apply at both the ``bulge'' mass scale as well as the BH mass scale.  The quantities $\delta$ and $\alpha$ are scales in space and time, respectively, and are related to the power-law index of the initial density perturbation $\epsilon$ in spherical infall models of halo growth (Henriksen \& Widrow 1999):
\begin{equation}
\frac{\delta}{\alpha}=\frac{2}{3}\left(1+\frac{1}{\epsilon}\right).
\label{eq:relate}
\end{equation}

This power-law index $\epsilon$ depends on the spectral index $n$ of the primordial power spectrum, $P(k) \propto k^n$, by $n = 2\epsilon - 3$.  There is therefore a direct link between the initial  power spectrum and the predicted relationship between the BH mass and the velocity dispersion; for agreement with observations, we must have $n = -2$, so that $\epsilon = 1/2$ and $\delta/\alpha = 2$.

We determine below that the addition of sufficient angular momentum breaks the self-similarity in the central regions.  This creates a ``mass reservoir'' around the BH that does grow in proportion to the galaxy mass, and from which the BH grows more slowly by collisional interactions between clumps of matter.  Provided most of this mass may be absorbed by the BH on a cosmic timescale, the proposed relation should still hold. 

\section{Numerical Simulations}

As an extension to the analytical work of MacMillan \& Henriksen (2002), we present here numerical simulations which confirm the predicted relationship between the BH mass and the velocity dispersion.

The initial conditions for the simulations follow those of Henriksen \& Widrow (1999): 
\begin{equation}
\rho(r, t_i) = \rho_c(t_i) [1 + \Delta(r, t_i)],
\end{equation}
where
\begin{equation}
\Delta(r, t_i) = \left\{ \begin{array}{ll} A\left[1 - B(r/r_c)^2\right] & r < r_c \\ A(1 - B)(r/r_c)^{-\epsilon}  & r \ge r_c. \end{array} \right.
\end{equation}
This initial density profile is that of a power-law function of radius, modified at small radii to account for suppression of small scale fluctuations.  The constant $B =  5 \epsilon/ (3 \epsilon + 6)$ and we choose $A = 0.5$.  The particles are started in the unperturbed Hubble flow.

Our initial numerical work began with a simple shell code, which is used to follow spherically symmetric shells of matter in an Einstein - de Sitter universe; each shell evolves according to
\begin{equation}
\frac{\mathrm{d} ^2 r_i}{\mathrm{d} t ^2} = -\frac{GM(r_i)}{r_i^2} + \frac{j^2}{r_i^3},
\end{equation}
where $M(r_i)$ is the total mass enclosed by the shell at radius $r_i$.  The angular momentum $j$ is given by $j^2 = 2 J G M(r_i) r_i$, evaluated at the initial shell positions.  Thus the particles are allowed to have, in general, elliptical orbits, and the shells follow the evolution of their radial coordinates.

To extend the spherically symmetric numerical results, we consider black hole growth in an n-body system of $n \sim 2000$ particles.  We use a modified version of Piet Hut's code (http://www.ids.ias.edu/~piet/act/comp/algorithms/starter/)
This code was modified to include a softening length to handle close encounters between particles.

In both simulations, a central BH has been included that can ``eat'' particles which fall below the Schwarzschild radius, $R_s = 2 G M_{\rm BH} / c^2$, and grow in mass accordingly.  Also, we choose a power-law index for the initial perturbation of $\epsilon = 1/2$, so that $\delta / \alpha = 2$ and $M_{\rm BH} \propto \sigma^4$, as observed.

\subsection{Results}

\begin{figure}
\centering
\includegraphics[scale=0.40]{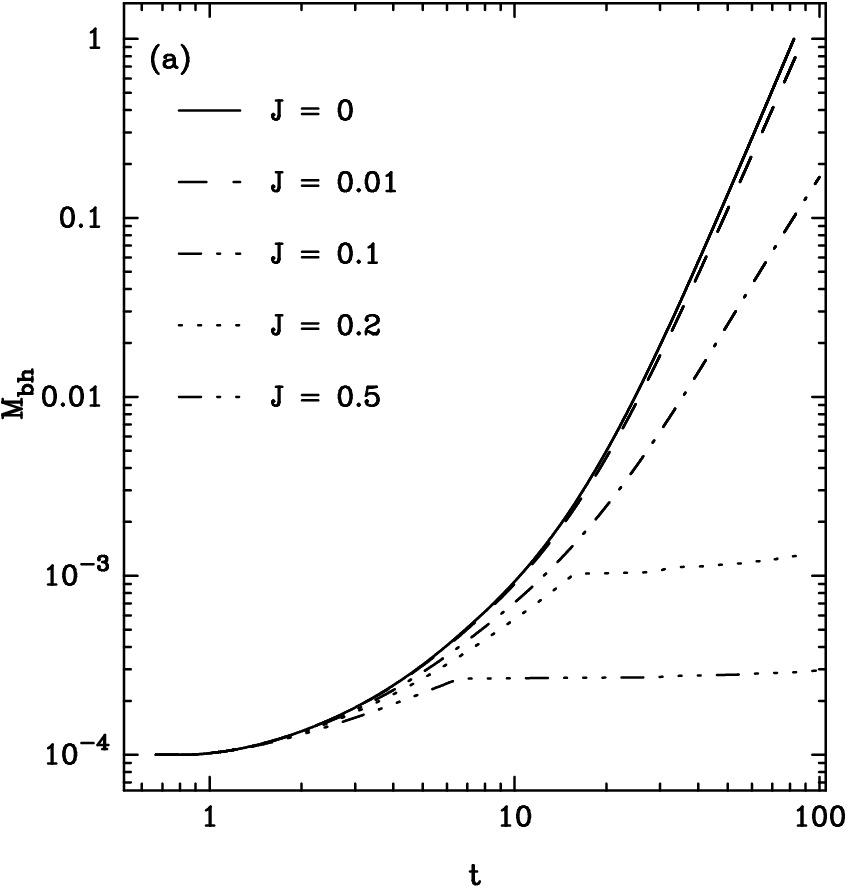}
\includegraphics[scale=0.40]{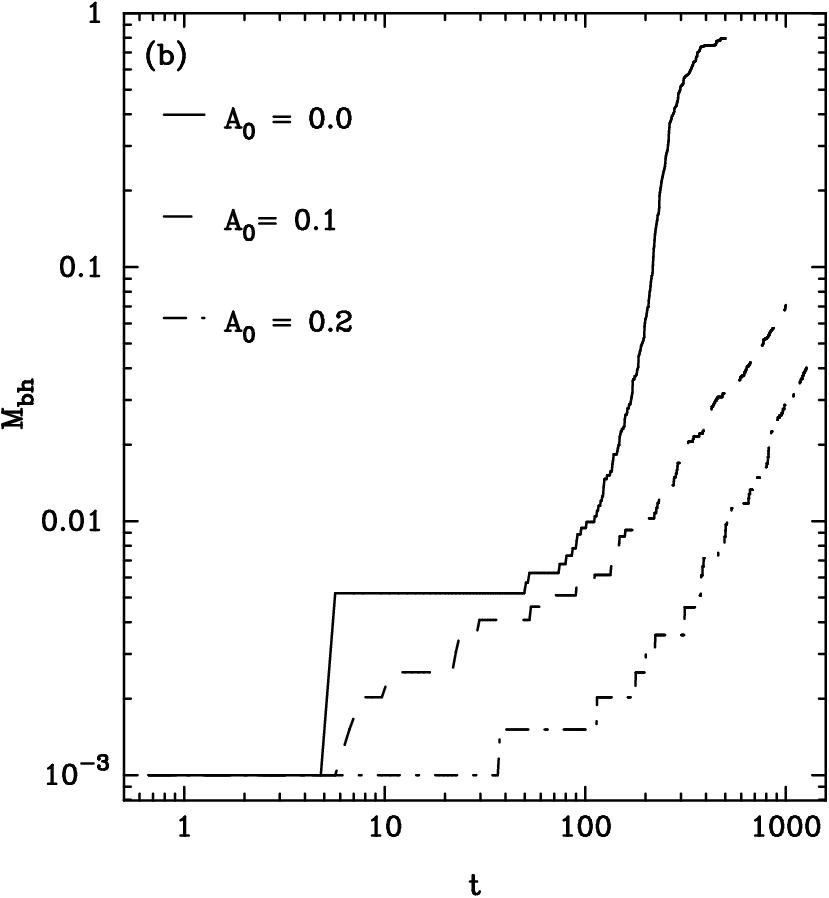}
\caption{Black hole growth as a function of time for (a) the shell code, and (b) the N-body code.}
\label{growth}
\end{figure}

The shell code simulations are run, with an initial BH mass of $10^{-4}$, until $t = 100$, at which time all matter is essentially ``eaten'' by the BH in the $J=0$ case.  Other values of the specific angular momentum are also explored.  The BH mass, as a function of time, is shown in Figure~\ref{growth}a for J = 0, 0.01, 0.1, 0.2, and 0.5.

MacMillan \& Henriksen (2002) predict that the BH mass should grow as $t^{4}$. This is indeed the asymptotic behavior for the case where $J=0$, as shown in the figure.  This logarithmic slope, however, decreases as the specific angular momentum increases; for $J = 0.1$, the mass is increasing as $t^{2.9}$.  For larger values of $J$, the growth essentially ``shuts off'' at some time, indicating that shells with high initial angular momentum are never reaching the centre of the system.

Three different cases are explored with the n-body simulations.  The first is an initial spherically symmetric system.  In the second and third cases, spherical symmetry is removed by giving each of the particles a random displacement, with amplitude $A_0$.  Systems are considered in which the amplitudes are about 0.8\% and 1.6\% of the initial extent of the system.

Results are shown in Figure~\ref{growth}b for the black hole mass as a function of time.  For the case with no initial random displacements ($A_0 = 0$), the growth is approximately that expected, although at later times the black hole growth slows down due to a lack of infalling particles.  For the more general cases where $A_0 > 0$, black hole growth remains self-similar, but with a much lower logarithmic slope: $M_{\rm BH} \propto t$ or a little greater.  In these systems, the particles have enough angular momentum that they don't fall straight into the central black hole, and rather form a ``core'' at the centre which slowly feeds the black hole growth.

\begin{figure}
\centering
\includegraphics[scale=0.40]{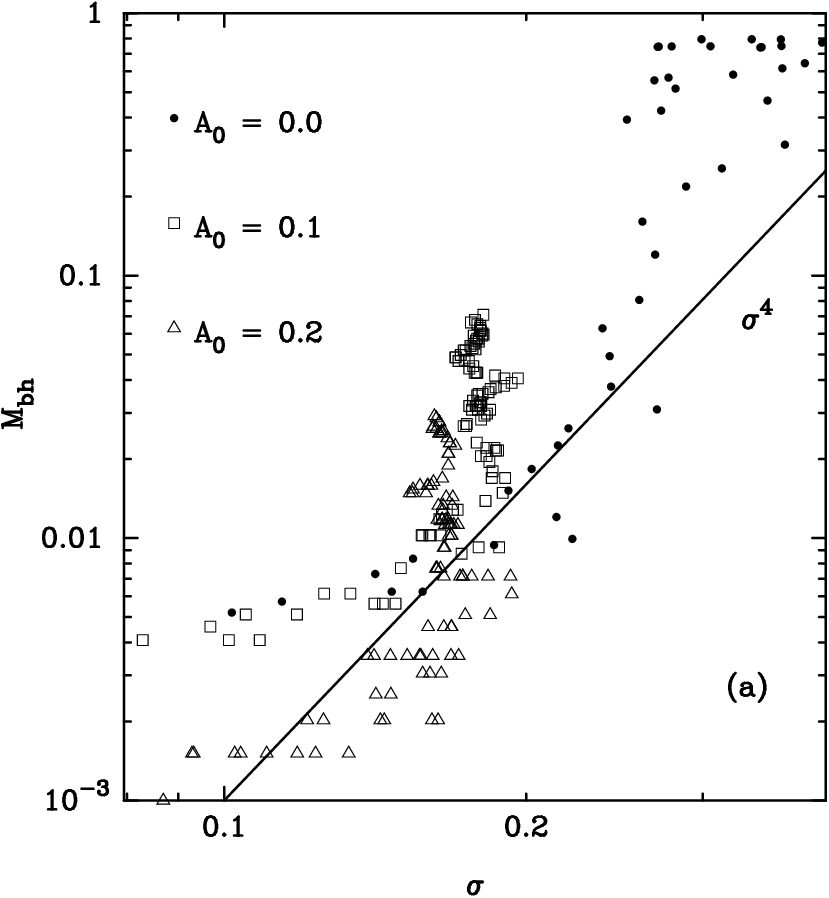}
\includegraphics[scale=0.40]{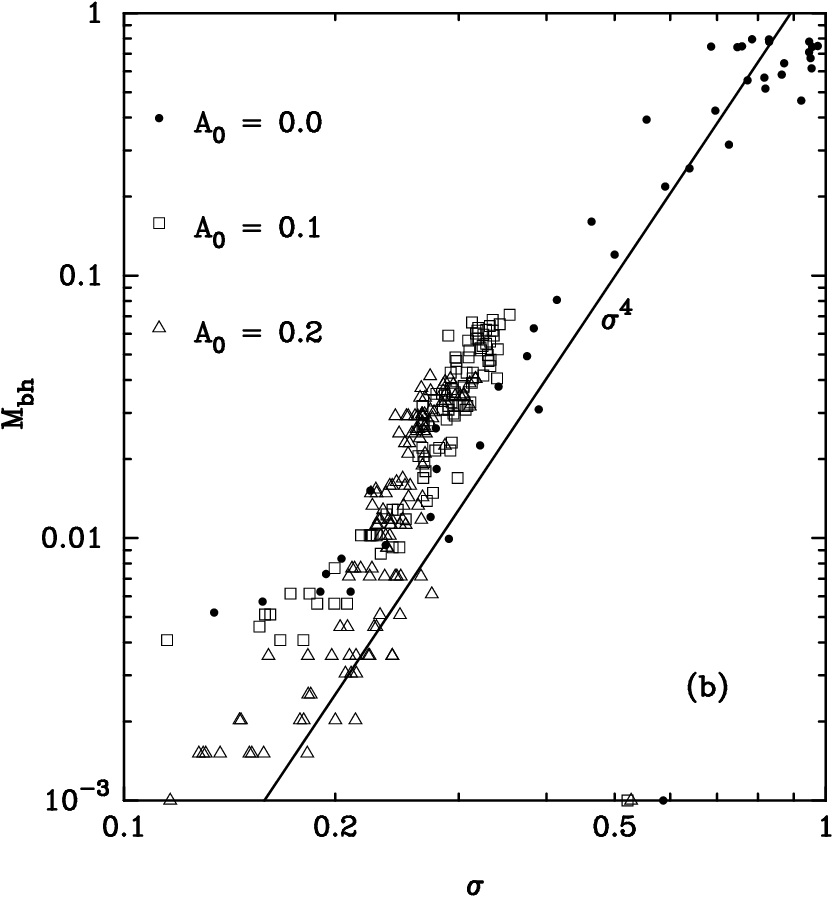}
\caption{The measured $M_{\rm BH} - \sigma$ relation for (a) a co-moving radius, and (b) a fixed radius.}
\label{msigma}
\end{figure}

To test the prediction that $M_{\rm BH} \propto \sigma^4$, the velocities of some fraction of the particles are averaged (equating $\sigma = \bar{v}$).  The  predictions  of MacMillan \& Henriksen (2002) apply to a fixed scaled position vector $\mathbf{X}$, implying that we should average all particles within some comoving radius.  Results are shown in Figure~\ref{msigma}a for a comoving radius of $r = 0.25$.  Although the case where $A_0 = 0$ may have a regime where it follows the predicted $M_{\rm BH}-\sigma$ relation, the cases with $A_0 > 0$ clearly follow no relation at all; their average velocity is about $0.18$ for most black hole masses.

Of interest, however, are the results if we take a fixed radius, rather than a comoving one, and average the velocities of the particles within that radius.  The plot of $M_{\rm BH}$ versus $\sigma$ is shown in Figure~\ref{msigma}b.  All three cases clearly show a trend that goes like $M_{\rm BH} \propto \sigma^4$.

\section{Discussion}

For systems in which there is no angular momentum or deviation from spherical symmetry initially, the numerical work, with both the spherical shell code and the more general n-body code, confirms the predicted black hole growth:  $M_{\rm BH} \propto t^4$.  However, for the simulations in which there was some deviation from a simple spherical infall of matter, results were contrary to what was expected.  In particular, the n-body results show a black hole growth approximately proportional to the time.  

However, it is apparent from the simulations that the black hole growth in these cases is not dominated by the infalling matter, as it is in the initially spherical case.  Rather, the particles form a ``core'' about the centre, and so the growth is fed in a different manner.  Dimensional arguments suggest that the Schwarzschild radius should scale with time as $R_s \propto ct$, where $c$ is the speed of light, a constant of the system.  If this is the case, then we get that the black hole mass will grow as $M_{\rm BH} \propto R_s \propto t$, which is approximately what is observed.  Note, however, that this breaks the symmetry of the system, since the mass on a larger scale is predicted to grow as $t^{3\delta - 2\alpha}$.

The numerically derived $M_{\rm BH} - \sigma$ relationship, which here compared the black hole growth with the average velocity of particles within some radius $r$, only gave the expected relation, $M_{\rm BH} \propto \sigma^4$, when a fixed, rather than comoving radius was considered.  Furthermore, the cases for which the growth rate was much shallower than expected still showed the same trend as the other case; that is, $M_{\rm BH} \propto \sigma^4$ regardless of how the black hole grew with time.

As stated above, however, this result is for a fixed radius.  From the definition of $\mathbf{X}$ and $T$ in MacMillan \& Henriksen (2002), we see that $r = \mathbf{X} t^{\delta / \alpha}$.  Fixing $r$ requires $\mathbf{X}$ to then depend on time.  Recalling that we're considering a density profile that goes as $X^{-\epsilon}$, and $\epsilon$ is related to the quantity $\delta / \alpha$ by equation (\ref{eq:relate}), we can write the mass as a function of $X$:
\begin{equation}
\mathcal{M}(X) \propto X^{3 - 2 \alpha / \delta}.
\end{equation}
Thus, as a function of the (fixed) radius, we have $M(r) \propto r^{3 - 2 \alpha / \delta}$.  Similarly, we can write the self-similar velocity, assuming it is in its virialized state, as a function of $X$:
\begin{equation}
Y(X) \propto \sqrt{\bar{\Phi}} \propto X^{1 - \alpha / \delta},
\end{equation}
so that the radial dependence of the velocity takes the form $v(r) \propto r^{1 - \alpha / \delta}$.

Combining these two equations for mass and velocity and eliminating $r$ gives us the familiar relation given by equation (\ref{eq:predict}).  Thus, taking the mass and velocity at a fixed radius $r$ gives the same relation regardless of the time dependence of either.

\section{Conclusions}

We consider black hole growth in a system which forms as the black hole is growing.  If this is the case, then the system has multidimensional self-similarity, and the black hole should grow as a power-law function of time, $M_{\rm BH} \propto t^{3\delta-2\alpha}$.  Furthermore, the velocities of the particles of the system will also evolve in a similar manner, and we predict that $\mathrm{log}(M_{\rm BH}) \propto (3\delta/\alpha-2)/(\delta/\alpha-1) \log{\sigma}$.

Numerically, we confirm that the black hole does indeed grow as a power-law function of time; however, it only follows the predicted relationship if the initial system has little or no angular momentum.  For systems with non-spherical initial conditions, the black hole grows as $M_{\rm BH} \propto t$ or a little greater.

Furthermore, we test the$M_{\rm BH} - \sigma$ relation with n-body simulations.  Although results do not show a strong relation if we take an averaged velocity within a comoving radius, the predicted relation, that $M_{\rm BH} \propto \sigma^4$, is clearly present if we take a fixed radius and average the particles within that.  This can be explained by considering the self-similar variable $\mathbf{X}$ to be a function of time such that the time dependence of the radial coordinate is fixed.

\end{document}